\documentclass[manuscript]{aastex}
\usepackage{emulateapj5}
\usepackage{times}
\usepackage{natbib}
\usepackage{epsfig}
\usepackage{rotating}
\usepackage{lscape}
\usepackage{apjfonts}
\usepackage{amssymb}

\def\arcs{\mbox{\ensuremath{^{\prime\prime}}}}
\def\farc{\mbox{\ensuremath{.\!\!\arcs}}}

\def\ga{\mathrel{\raise0.35ex\hbox{$\scriptstyle >$}\kern-0.6em
\lower0.40ex\hbox{{$\scriptstyle \sim$}}}}
\def\la{\mathrel{\raise0.35ex\hbox{$\scriptstyle <$}\kern-0.6em
\lower0.40ex\hbox{{$\scriptstyle \sim$}}}}
\def\co{CO~{\it J}=1-0 }
\def\kms{km~s$^{-1}$}

\def\cii{C{\sc ii}-158$\mu$m}

\def\cothree{CO~{\it J}=3-2 }
\def\hcm{HCM\,6A}

\def\arcs{\hbox{$^{\prime\prime}$}}

\shorttitle{Search for C{\sc ii}-158$\mu$m line emission in HCM\,6A}
\shortauthors{Kanekar et al.}
\slugcomment{ApJL draft}

\begin{document}

\title{A search for C{\sc ii}-158$\mu$m line emission in HCM\,6A, a Lyman-$\alpha$ emitter at $z=6.56$}

\author{Nissim~Kanekar\altaffilmark{1}, 
Jeff~Wagg\altaffilmark{2,3}, 
Ranga Ram Chary\altaffilmark{4},
Christopher~L.~Carilli\altaffilmark{5}}

\altaffiltext{1}{Ramanujan Fellow; National Centre for Radio Astrophysics, Tata 
Institute of Fundamental Research, Pune 411 007, India; nkanekar@ncra.tifr.res.in}
\altaffiltext{2}{European Southern Observatory, Alonso de C\'ordova 3107, 
Vitacura, Casilla 19001, Santiago 19, Chile}
\altaffiltext{3}{Astrophysics Group, Cavendish Laboratory, University of Cambridge, Cambridge, CB3 0HE, UK}
\altaffiltext{4}{U.S. Planck Data Center, MS220-6 Caltech, Pasadena, CA 91125, USA}
\altaffiltext{5}{National Radio Astronomy Observatory, PO Box O, Socorro, NM 87801, USA}

\begin{abstract}
We report a Plateau de Bure interferometer search for C{\sc ii}-158$\mu$m emission from 
HCM\,6A, a lensed 
Lyman-$\alpha$ emitter (LAE) at $z = 6.56$. Our non-detections of C{\sc ii}-158$\mu$m
line emission and 1.2\,mm radio continuum emission yield $3\sigma$ limits of 
L$_{\rm CII} < 6.4 \times 10^7 \times (\Delta V/100 km s^{-1})^{1/2}$~L$_\odot$ 
on the C{\sc ii}-158$\mu$m line luminosity and S$_{\rm 1.2mm} < 0.68$~mJy on the 
1.2\,mm flux density. The local conversion factor between L$_{\rm CII}$ and star 
formation rate (SFR) yields an SFR~$< 4.7$~M$_\odot$~yr$^{-1}$, $\approx 2$ times lower 
than that inferred from the ultraviolet (UV) continuum, suggesting that the local factor 
may not be applicable in high-$z$ LAEs. The non-detection of 1.2\,mm continuum emission 
yields a total SFR~$< 28$~M$_\odot$~yr$^{-1}$; any obscured star formation is thus within a factor 
of two of the visible star formation. Our best-fit model to the rest-frame UV/optical 
spectral energy distribution of HCM\,6A yields a stellar mass of $1.3 \times 10^9$\,M$_\odot$ 
and an SFR of $\approx 10$\,M$_\odot$~yr$^{-1}$, with negligible dust obscuration. We 
fortuitously detect CO~J$=3-2$ 
emission from a $z = 0.375$ galaxy in the foreground cluster Abell\,370, obtaining a 
CO line luminosity of L$^\prime ({\rm CO}) > (8.95 \pm 0.79) \times 10^8$~K~km~s$^{-1}$~pc$^2$, 
and a molecular gas mass of M$({\rm H_2}) > (4.12 \pm 0.36) \times 10^9$~M$_\odot$, for a 
CO-to-H$_2$ conversion factor of 4.6~M$_\odot$~(K~km~s$^{-1}$~pc$^2$)$^{-1}$.
\end{abstract}

\keywords{cosmology: observations --- galaxies: evolution --- galaxies: formation --- infrared: galaxies}


\section{Introduction} 


A large population of Lyman-$\alpha$ emitters (LAEs) have recently been detected at 
$z \gtrsim 6$, towards the end of the epoch of reionization 
\citep[e.g.][]{hu02,taniguchi05,kashikawa11}. LAEs appear to be normal star-forming 
galaxies, with star-formation rates (SFRs) of $\approx 5 - 60$~M$_\odot$~yr$^{-1}$.
Their space density is sufficiently high that low-luminosity LAEs may be the primary 
source of the UV photon background required for reionization \citep{kashikawa11}.
Detailed studies of the high-$z$ LAE population are hence of much interest. 

At high redshifts, $z > 6$, and for massive galaxies and quasars, radio CO lines have 
proved good tracers of the gas associated with star formation 
\citep[e.g.][]{walter03,wang10,wang11}. Unfortunately, deep CO searches in LAEs at 
similar redshifts ($z \sim 6.5-7$) have so far only yielded non-detections 
\citep{wagg09,wagg12a}.


The \cii\ $^2P_{3/2} \rightarrow {^2}P_{1/2}$ transition is one of the brightest lines in 
the spectrum of any galaxy, often carrying $\approx 0.5$\% of the total galaxy luminosity 
\citep[e.g.][]{crawford85,stacey91}. It is the primary coolant of the diffuse interstellar 
medium at temperatures $< 5000$~K \citep[e.g.][]{wolfire95} and thus an excellent tracer of 
cold neutral gas, the fuel for star formation. While \cii\ emission has long been known to 
arise from photo-dissociation regions \citep[e.g.][]{crawford85}, the \cii\ emission from 
extended diffuse gas in normal galaxies has been recently shown to be comparable to that 
from dense star-forming regions \citep{pierini01}. \cii\ emission studies 
may thus also provide information on the kinematics and dynamical masses of high-$z$ 
galaxies, besides the spatial distribution of star-forming gas. 

The \cii\ transition thus provides a useful tracer of physical conditions in $z > 6$ 
star-forming galaxies \citep[e.g.][]{carilli13}. While \cii\ emission has been found in a 
number of star-forming galaxies at $z \approx 1.2$ \citep{stacey10}, most high-$z$ 
($z \gtrsim 4$) searches have targeted massive objects with a high far-infrared 
(FIR) luminosity, L$_{\rm FIR} > 10^{12}$~L$_\odot$ 
\citep[e.g.][]{maiolino05,maiolino09,wagg10,debreuck11}. Recently, the first searches 
for \cii\ emission in LAEs were reported by \citet{walter12a}. In this {\it Letter}, we 
report results from a deep search for \cii\ emission in a lensed LAE at $z = 6.56$.

\section{HCM\,6A: A lensed LAE at $z \sim 6.56$}
\label{sec:source}


\hcm\ was identified by \citet{hu02} as an LAE at $z = 6.56$ in the field of the $z = 0.375$ 
galaxy cluster Abell\,370. A lensing model for Abell\,370 yields a magnification of $\approx 4.5$ 
at the LAE location \citep{kneib93}. \citet{hu02} obtained an SFR of $\sim 9$~M$_\odot$~yr$^{-1}$ 
\citep[see][]{hu02err} from the rest-frame UV continuum, a value typical of the $z \gtrsim 6$ 
LAE population \citep[e.g.][]{taniguchi05,cowie11}. However, a far higher SFR, 
$\sim 140$~M$_\odot$~yr$^{-1}$, was inferred by \citet{chary05} from the 
excess emission detected in the {\it Spitzer} 4.5~$\mu$m image of the field, suggesting 
significant dust extinction of the UV continuum. A high SFR, $11-41$~M$_\odot$~yr$^{-1}$, was 
also obtained by \citet{schaerer05}, from modelling the spectral energy distribution (SED). 
Conversely, \citet{boone07} used a 1.2\,mm MAMBO-2 image to obtain a strong constraint on the 
far-infrared (FIR) luminosity, and thence on the SFR, $< 35$~M$_\odot$~yr$^{-1}$. The high 
magnification factor and the possible high SFR makes \hcm\ an excellent target for a 
search for \cii\ emission from a high-$z$ LAE.

\section{Observations, data analysis and results}
\label{sec:obs}

\begin{figure*}
\centering
\includegraphics[scale=0.45]{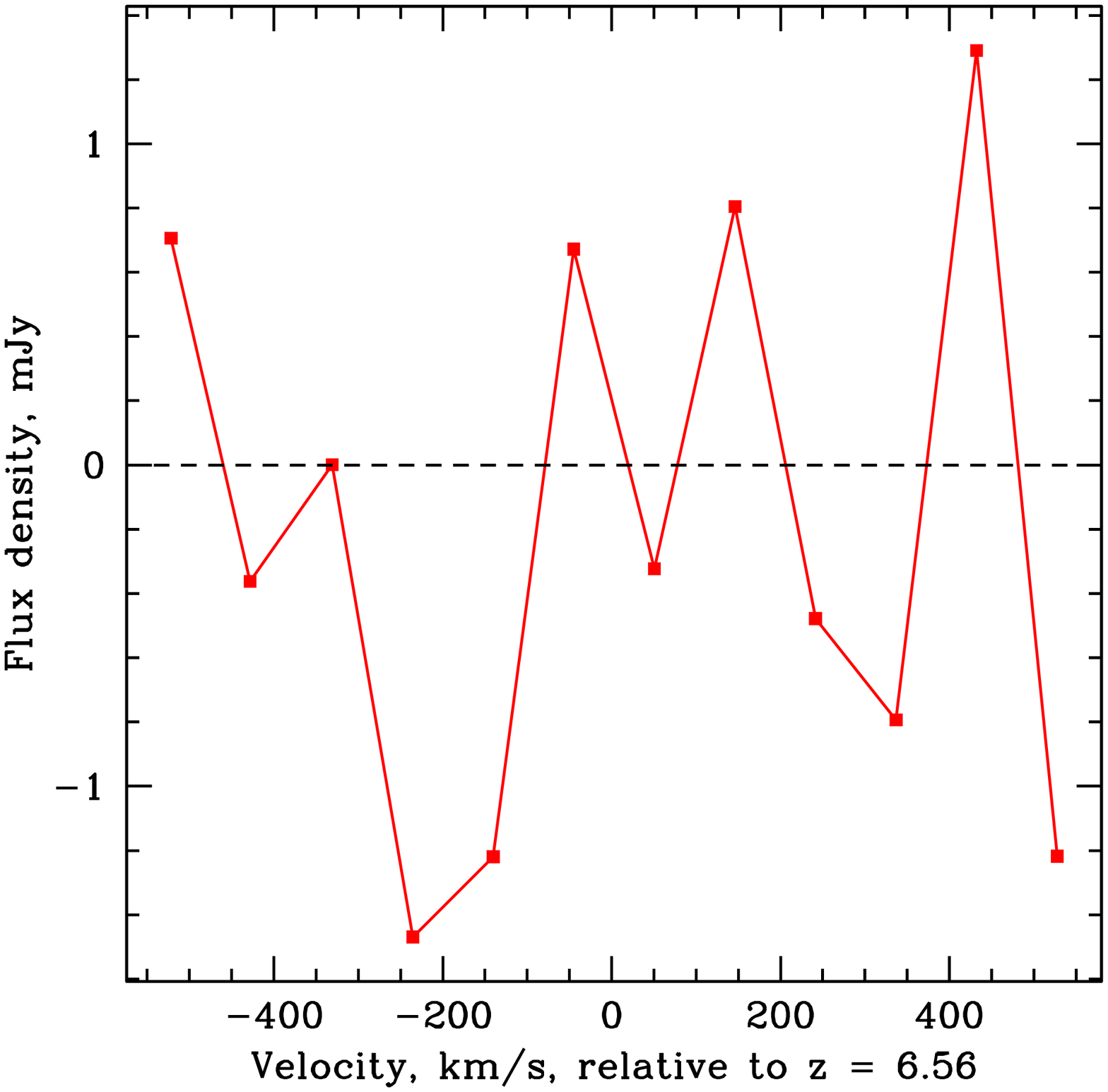}
\includegraphics[scale=0.45]{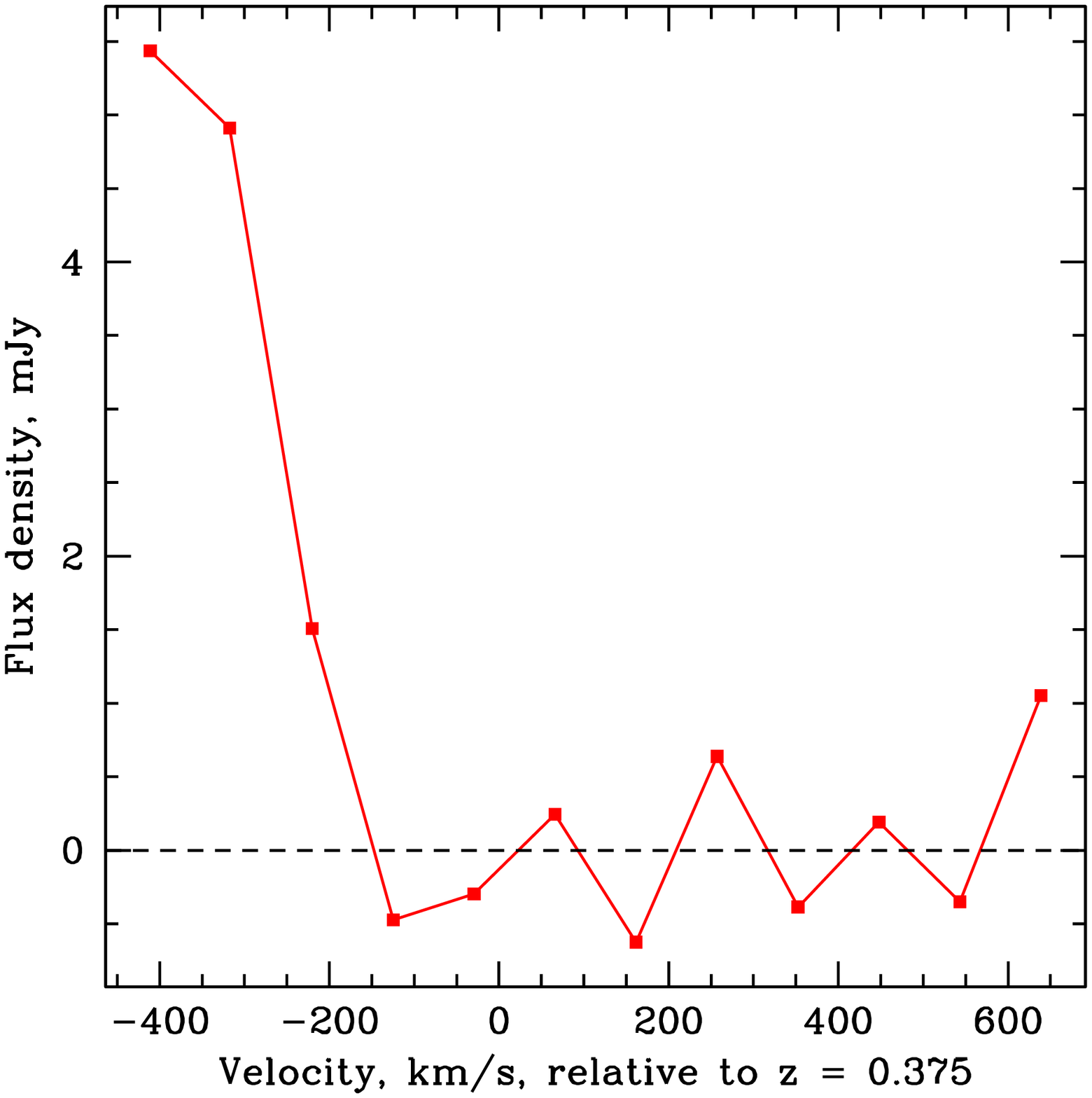}
\caption{Left panel: The spectrum towards \hcm, covering the \cii\ line; the velocity scale 
is relative to the redshifted \cii\ line frequency from $z = 6.56$.  Right panel: The 
spectrum obtained from the bright spiral galaxy $\sim 5\farc$ southwest of \hcm; the 
velocity scale is relative to the redshifted \cothree\ line frequency from $z = 0.375$.}
\label{fig:spectra}
\end{figure*}

We searched for redshifted \cii\ line emission from \hcm\ using five telescopes of the Plateau 
de Bure Interferometer (PdBI) on August 18 and 21, 2009, with the array in the compact 5Dq 
configuration. The observations used a bandwidth of 1~GHz, centred at the redshifted \cii\ 
line frequency of 251.395~GHz, with two orthogonal polarizations and 256 channels. 
The velocity coverage was $\approx 1200$~\kms, with a velocity resolution of $\approx 4.7$~\kms.
This coverage is sufficient to cover errors in the Ly$\alpha$ emission redshift due to 
Ly$\alpha$ absorption in the IGM, as well as scenarios where the Ly$\alpha$ emission arises 
from outflowing gas (typical velocities of a few hundred km~s$^{-1}$ for small galaxies like LAEs). 

The complex antenna gains were calibrated on the nearby quasars 0235+164 and 0336$-$019, the 
antenna bandpass shapes measured on 3C454.3, and the flux density scale derived from observations 
of MWC349. The total observing time was 18~hours, in two full-synthesis tracks, with an on-source 
integration time of 10.5~hours. System temperatures were lower than 200~K during most of the run.

The data were analysed using standard calibration procedures in the 
{\sc GILDAS}\footnote{http://www.iram.fr/IRAMFR/GILDAS} package, followed by imaging in the 
{\sc AIPS} package. Line-free channels were averaged together to produce a ``channel-0'' dataset 
for the purpose of continuum imaging; our final continuum map has an angular resolution of 
$\sim 2\farc 5 \times 1\farc 7$ and a root-mean-square (RMS) noise of 0.16~mJy/Bm. No radio 
continuum emission was detected in the field.


\begin{figure*}[t!]
\centering
\includegraphics[scale=0.45]{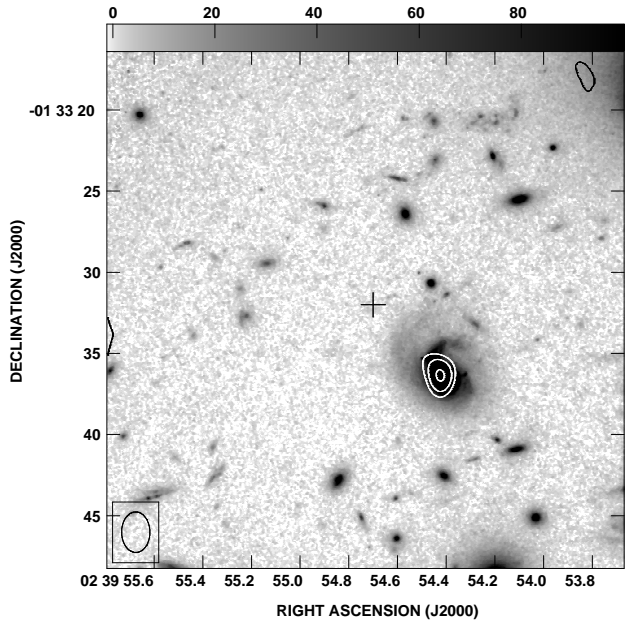}
\caption{An overlay of the integrated \cothree\ emission (in contours) on the HST-ACS optical 
image (in greyscale). The emission feature arises from the bright spiral, $5\farc$ southwest 
of \hcm, whose position is marked by a ``+''.}
\label{fig:tst}
\end{figure*}

The final image cube was made using ``natural'' weighting, to maximize the sensitivity, 
after Hanning-smoothing the visibility data, and resampling to a velocity resolution of 
100~\kms. No line emission was detected at the location of \hcm\ in the cube, but a strong 
emission feature was detected $\sim 5\farc$ south-west of \hcm. This was identified as 
redshifted \cothree\ emission from a galaxy in the foreground cluster, at $z = 0.375$ 
(see below). The spectrum obtained at the location of \hcm\ in the CLEAN'ed image cube 
is shown in 
the left panel of Fig.~\ref{fig:spectra}, after smoothing to, and resampling at, a resolution 
of $\sim 100$~\kms.  It has an RMS noise of 0.6~mJy per 100~\kms\ channel (measured in the image plane). 

The optical image of \hcm\ is elongated along a position angle of 110$^\circ$, with a 
length of $\sim 4\farc$ \citep{hu02}. The extended structure is covered by two of our 
synthesized beams; our limits on the total \cii\ luminosity and the 1.2\,mm continuum 
flux density are hence worse by a factor of $\approx \sqrt{2}$ than the limits from 
a single beam. The non-detections of \cii\ line emission and 1.2\,mm radio continuum 
emission from \hcm\ then yield the $3\sigma$ upper limits 
L$_{\rm [CII]} < 6.4 \times 10^7 \times (\Delta V/100)^{1/2}$~L$_\odot$ on the \cii\ line 
luminosity (assuming a line FWHM of 100~\kms, typical for small galaxies)
and $S_{\rm 1.2mm} < 0.68$~mJy, after correcting the former for the magnification 
factor of 4.5.\footnote{We use a Lambda-CDM model with $H_0 = 71$~km~s$^{-1}$~Mpc$^{-1}$, 
$\Omega_{\rm m} = 0.27$ and $\Omega_{\Lambda} = 0.73$ 
\citep{spergel07}.}

We also revisited the {\it Spitzer} photometry of \hcm\ \citep{chary05}. Since those results, 
which used data from {\it Spitzer} program 64 with an integration time of 2400s, the field 
has been observed four times in programs 137 and 60034. We have re-analyzed all these data,
clearly detecting \hcm\ in each epoch. The total exposure time corresponds to more than 10~hrs 
of imaging. 

We obtained the post-Basic Calibrated Data mosaics of the \hcm\ field for each program from the 
{\it Spitzer} Heritage Archive. The astrometric alignment of each mosaic was assessed through 
bright sources. Of the three programs, the astrometric alignment of program 60034 appears to 
be $1.5-2$ times worse than the others, with a median offset of $0\farc4$ relative to 
the location of the same sources in the Sloan Digital Sky Survey images. Given the high 
signal-to-noise ratio of the source in each of the mosaics, we simply ignore the frames with 
worse astrometry rather than applying an astrometric correction to those frames.

The $3.6 \mu$m and $4.5\mu$m photometry of \hcm\ was carried out both in the mosaics from the 
individual {\it Spitzer} observations, as well as from a super-mosaic of all the data with 
good astrometry. The super-mosaic was constructed using {\sc swarp}, via an 
integration-time-weighted median combination of the mosaics with good astrometry.

Due to the contamination of the flux density of \hcm\ by the bright extended spiral galaxy 
5$\arcsec$ to the south-west, we used {\sc galfit} to model the spatial profile of the 
spiral and subtract it out. HCM\,6A's flux density in the different mosaics spans the range 
$0.3-0.6$~$\mu$Jy at $3.6\mu$m, and $1-1.4$~$\mu$Jy at $4.5\mu$m, with systematics from the 
subtraction of the spiral dominating the errors. The statistical uncertainty on the 
flux density values is $0.1 \: \mu$Jy.

\begin{figure*}
\centering
\includegraphics[scale=0.45]{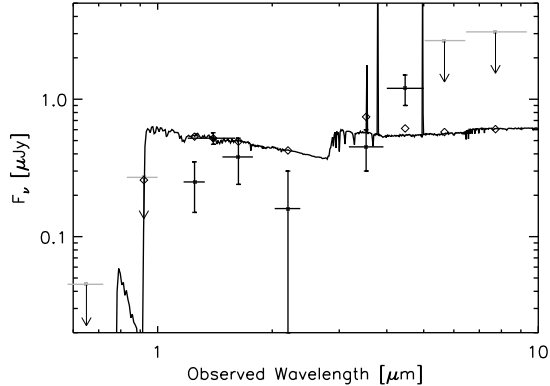}
\caption{Fit to the SED of \hcm\ at rest-frame UV to optical wavelengths, with a modified 
sub-solar ($Z = 0.2 Z_\odot$) metallicity BC03 model. The points with error bars are the various 
measurements, while the open diamonds indicate values from the best fit model. See text for 
discussion.}
\label{fig:sed}
\end{figure*}

We used a modified version of the \citet{bruzual03} (BC03) population synthesis models to 
fit the complete SED of \hcm\ over the rest-frame UV to optical wavelength range,
using our {\it Spitzer} measurements, the recent $1.4 \mu$m detection \citep{cowie11}, 
and the data of \citet{hu02}. Our modification of the BC03 model involved the inclusion of 
nebular line emission, particularly the energetically important Ly$\alpha$, [OII], H$\beta$, 
[O{\sc iii}], H$\alpha$ and [N{\sc ii}] lines. The relative line ratios correspond to those 
observed in nearby H$\alpha$ emitters, which have been argued to be local analogs of 
high-$z$ star-forming galaxies \citep{shim13}. The absolute line strengths were calibrated 
with respect to the FUV continuum at 1500\AA\ derived from the BC03 models. Since nebular line
emission is correlated with Lyman-continuum emission and not FUV continuum emission, this 
is potentially an inaccurate assumption. We added these emission lines to the BC03 templates 
for stellar photospheric emission and then fit the multi-wavelength photometry, taking the 
IGM absorption into account. The high signal-to-noise of the photometry in \citet{cowie11} 
drives the best-fit model, shown in Fig.~\ref{fig:sed}; this does a poor job of reproducing 
the {\it Spitzer} photometry. While the excess emission in the 4.5$\mu$m band suggests 
H$\alpha$ contamination \citep{chary05}, the {\it Spitzer} photometry is affected by 
systematics due to source confusion with the foreground spiral. Our best-fit SED model 
yields a stellar mass of $1.3 \times 10^9$~M$_\odot$ and an SFR of 
$\approx 10$~M$_\odot$\,yr$^{-1}$ for \hcm, with negligible dust obscuration. 


The \cothree\ spectrum obtained towards the $z = 0.375$ galaxy in Abell\,370 is shown 
in the right panel of Fig.~\ref{fig:spectra}, again at a resolution of $\sim 100$~\kms, but 
with the velocity scale relative to the redshifted \cothree\ line frequency at $z = 0.375$. 
Fig.~\ref{fig:tst} shows an overlay of the integrated emission (in contours) on an 
optical image from the Advanced Camera for Surveys (ACS) onboard the {\it Hubble Space Telescope} 
(HST) in the F814W band \citep{richard10}. The emission feature clearly arises from the bright 
cluster galaxy $5\farc$ to the southwest of 
\hcm. Unfortunately, the feature is right at the edge of our observing band, with part of the 
emission outside the band. We hence obtain a {\it lower limit} to the line flux of 
$1.13 \pm 0.10$~Jy~\kms, after ``CLEAN''ing the cube. Assuming that the \cothree\ line is 
thermalized, this implies a CO line luminosity of L$^\prime ({\rm CO}) > (8.95 \pm 0.79) 
\times 10^8$~K~km~s$^{-1}$~pc$^2$.

\section{Discussion}
\label{sec:discuss}

It has been suggested that the \cii\ line luminosity might be used to estimate 
the SFR in star-forming galaxies \citep[e.g.][]{stacey91,boselli02,delooze11}, 
unbiased by dust extinction. \citet{delooze11} found a good correlation between the 
\cii\ line luminosity and the SFR (determined from a combination of the far-UV 
continuum and the 24~$\mu$m continuum magnitudes) for 24 nearby star-forming galaxies, 
with Log[SFR$_{{\rm [FUV + 24}\mu{\rm m]}} /({\rm M}_\odot\:{\rm yr}^{-1})$]~$= 
0.983 \times {\rm Log[L}_{\rm [CII]}/({\rm erg\: s^{-1}}){\rm ]} - 40.012$, 
The relation, which has a $1\sigma$ dispersion of $0.27$~dex, is expected 
to be applicable to star-forming galaxies with SFR~$ \sim 0.05 - 127$~M$_\odot$~yr${-1}$.

Our non-detection of \cii\ emission in \hcm\ gives L$_{\rm [CII]} < 6.4 \times 10^7
(\Delta V/100)^{1/2}$~L$_\odot$. This is a significantly tighter constraint than obtained 
by \citet{walter12a} in two LAEs at $z \approx 6.6-7.0$, mostly due to the large 
magnification factor in \hcm. Using the relation of \citet{delooze11}, this implies an upper 
limit of $\sim 4.7$~M$_\odot$~yr$^{-1}$ to the total SFR (both obscured and unobscured). However, 
the lower limit to the SFR from the UV continuum is $\sim 9$~M$_\odot$~yr$^{-1}$, 
a factor of $\sim 2$ higher than the above upper limit. Further, the discrepancy between 
the predicted and observed SFRs would be exacerbated if dust extinction is indeed 
significant in the LAE \citep{chary05,schaerer05}. Our results thus suggest that the 
local relation between \cii\ line luminosity and SFR may not be applicable to 
high-$z$ LAEs (although we note that the dispersion in the local relation is a factor of 
$\approx 1.9$).

Our upper limit to the 1.2\,mm radio continuum flux density of \hcm\ ($0.68$\,mJy)
is slightly tighter than that of \citet{boone07} ($1.08$\,mJy). Since the observing 
frequency of 1.2\,mm corresponds to a rest frequency of $158.6 \mu$m at $z = 6.56$, we will 
use the conversion factor of \citet{calzetti10} (valid for star-forming and starburst galaxies) 
to convert from the $160\mu$m luminosity to an SFR. We obtain a $3\sigma$ upper limit of 
$28$~M$_\odot$~yr$^{-1}$ on the SFR, after including the lensing factor of 4.5. 
This is within a factor of 3 of the SFR derived from the UV continuum \citep{hu02}, 
indicating that any obscured star formation in \hcm\ is at most two times larger than 
the visible star formation. Using the same assumptions as \citet{boone07} yields 
a $3\sigma$ SFR upper limit of $22$~M$_\odot$~yr$^{-1}$, and does not significantly 
change the above result.

Our best-fit SED model yields an SFR of $\approx 10$~M$_\odot$~yr$^{-1}$, consistent
with the SFR estimates from the UV continuum and the upper limit to the 1.2\,mm continuum
flux density. \citet{chary05} inferred a far higher SFR ($140$~M$_\odot$~yr$^{-1}$) from 
the excess 4.5\,$\mu$m emission, assuming this to arise from H$\alpha$ contamination. However, 
the SFR derived from the H$\alpha$ line flux critically depends on the age of the stellar 
population. If \hcm\ is indeed similar in physical properties to the $z\sim 5$ H$\alpha$ 
emitters reported in \citet{shim11}, and the local H$\alpha$ emitters of \citet{shim13}, 
the photometry is dominated by a population of massive O stars which excite H$\alpha$ 
emission due to their strong ionizing photon field. In such a situation, the same H$\alpha$ 
line flux corresponds to a {\it lower} SFR than that obtained from the standard calibration 
\citep{kennicutt98}. Thus, depending on the age of the stellar population, the same 
H$\alpha$ line flux could correspond to a SFR in the range $30-150$~M$_{\odot}$/yr. 

\begin{figure*}[t!]
\centering
\includegraphics[scale=0.5]{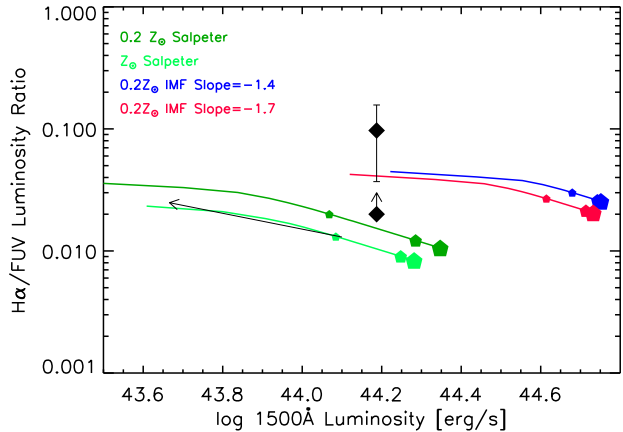}
\caption{The temporal evolution of the ratio of H$\alpha$ to far-UV flux for different 
stellar populations, evaluated using {\it Starburst99}; the stars of increasing size are
at 0.01, 0.1 and 1~Gyr, and the arrow shows 1 magnitude of extinction at 1500\,\AA. The 
lowest two curves are for Salpeter stellar populations with solar and 0.2~solar 
metallicity, respectively, while the top two curves are for populations with 
0.2~solar metallicity, but IMF slopes of $-1.4$ and $-1.7$. The two diamonds indicate
the value of the H$\alpha$/FUV flux ratio in \hcm, the upper one giving the measured ratio, and 
the lower one accounting for possible dust obscuration of the FUV continuum (see main text). }
\label{fig:hauv}
\end{figure*}

The ratio of a galaxy's luminosity in the H$\alpha$ and FUV bands is a sensitive tracer of 
the properties of the stellar population, including its metallicity, initial mass function 
(IMF) and age, and the extent of dust obscuration \citep[e.g.][]{kennicutt98,shim11}. 
We note that the SFR estimates from the UV continuum and the H$\alpha$ line are only valid for a 
Salpeter IMF and a 100-Myr-old continuously star-forming stellar population. 
At low ages ($<10$~Myr), there are more Lyman-continuum and UV photons produced per unit 
baryon in stars; a lower SFR can thus yield a higher H$\alpha$/UV flux ratio.
Comparing the observed ratio in high-$z$ LAEs with predictions from population synthesis 
models thus provides insights on early stellar populations. We have investigated 
this via a simulation in {\it Starburst99} \citep{leitherer99}, 
using constant star-formation models (SFR~$= 1$~M$_\odot$~yr$^{-1}$) with a range of 
metallicities ($0.02<Z/Z_{\sun}<1$) and IMFs. In \hcm, the measured H$\alpha$/FUV ratio 
is $\approx 0.03-0.16$, including systematic errors in the H$\alpha$ luminosity; this 
is consistent with values expected from low-metallicity young stellar populations with a 
top-heavy IMF (see Fig.~\ref{fig:hauv}). However, the intrinsic ratio could be lower 
than this if dust preferentially obscures the FUV continuum. Combining our upper limit 
on the 1.2\,mm flux density with the FUV photometry yields a lower limit of $0.02$ on 
the H$\alpha$/FUV ratio, and thus an allowed range of $0.02-0.16$, marginally consistent 
with values expected from standard stellar populations and IMFs at low metallicity. 
Note that increasing the SFR would cause the four curves to shift towards the right. 
While a young ($< 1$~Myr) starburst can explain the observed ratio, such an age is unlikely 
due to the low probability of detecting such a young starburst at a time when the Universe 
is $\gtrsim 1$~Gyr old.

For the galaxy detected in \cothree\ emission, assuming a CO-to-H$_2$ conversion factor
allows us to infer its molecular gas mass. The $z = 0.375$ galaxy has been classified 
as a spiral by \citet{mellier88}, based on its optical colours. We hence use the Milky Way 
conversion factor $X_{\rm CO} = 4.6$~M$_\odot$~(K~\kms~pc$^{-2}$)$^{-1}$ \citep{solomon91} to 
obtain M$({\rm H_2}) > (4.12 \pm 0.36) \times 10^9$~M$_\odot$. This assumes that the 
\cothree\ transition is thermalized. However, \citet{dannerbauer09} find that large spiral 
galaxies at $z \sim 1.5$ have sub-thermal \cothree\ line emission, similar to the situation in 
the Galaxy \citep[e.g.][]{fixsen99}.  Our estimate thus provides a lower limit to the 
molecular gas mass because (1)~the \cothree\ emission may be sub-thermal, and 
(2)~part of the CO emission lies outside our observing band. Finally, we do not 
spatially resolve the CO emission, indicating that it arises from a region smaller than 
$\sim 9$~kpc~$\times 13$~kpc.  

The $z = 0.375$ galaxy (BO~39) is located $\approx 330$~kpc from the cluster centre, well 
within the cluster virial radius ($\approx 2.6$~Mpc); it has a velocity of $+100$~\kms\ relative 
to $z = 0.374$ \citep{mellier88}. The detection of CO emission from a galaxy close to 
the centre of a large cluster allows a test of whether 
ram pressure stripping can denude a galaxy of its molecular gas, as is believed to 
occur for the atomic gas \citep[e.g.][]{gunn72,chung07}. We note that the CO line 
emission peaks at $z \approx 0.373$, while the galaxy redshift is $z = 0.3745$ 
\citep{mellier88}. Follow-up mapping \co\ studies will be of much interest.

\section{Summary}
\label{sec:sum}

We report strong constraints on the \cii\ line luminosity and the 1.2\,mm radio 
continuum flux density of the $z = 6.56$ LAE, \hcm, from a PdBI imaging study. 
The \cii\ non-detection suggests that the local conversion factor from \cii\ line luminosity 
to SFR may not be applicable in high-$z$ galaxies. Our upper limit to the SFR from the 
1.2\,mm continuum non-detection indicates that the LAE does not have significant obscured star 
formation. Our best-fit model to the SED of \hcm\ yields a stellar mass of $1.3\times 10^9$~M$_\odot$ 
and an SFR of $\approx 10$~M$_\odot$~yr$^{-1}$, with negligible dust obscuration. This 
is consistent with the SFR estimates from the UV continuum and the 1.2\,mm radio continuum. 
We detect \cothree\ emission from a $z=0.375$ spiral galaxy in the foreground cluster 
Abell\,370, and obtain the lower limit M$({\rm H_2}) > (4.12 \pm 0.36) \times 10^9$~M$_\odot$ 
on its molecular gas mass.

\section*{Acknowledgments}

NK, JW and CC are grateful for support from the Max-Planck Society and the 
Alexander von Humboldt Foundation. NK acknowledges support from the 
Department of Science and Technology through a Ramanujan Fellowship. 
The National Radio Astronomy Observatory is a facility of the National 
Science Foundation operated under cooperative agreement by Associated Universities, Inc.

\bibliographystyle{apj}

\begin{thebibliography}{41}
\expandafter\ifx\csname natexlab\endcsname\relax\def\natexlab#1{#1}\fi

\bibitem[{{Boone} {et~al.}(2007){Boone}, {Schaerer}, {Pell{\'o}}, {Combes}, \&
  {Egami}}]{boone07}
{Boone}, F., {Schaerer}, D., {Pell{\'o}}, R., {Combes}, F., \& {Egami}, E.
  2007, A\&A, 475, 513

\bibitem[{{Boselli} {et~al.}(2002){Boselli}, {Gavazzi}, {Lequeux}, \&
  {Pierini}}]{boselli02}
{Boselli}, A., {Gavazzi}, G., {Lequeux}, J., \& {Pierini}, D. 2002, A\&A, 385,
  454

\bibitem[{{Bruzual} \& {Charlot}(2003)}]{bruzual03}
{Bruzual}, G. \& {Charlot}, S. 2003, MNRAS, 344, 1000

\bibitem[{{Calzetti} {et~al.}(2010){Calzetti}, {Wu}, {Hong}, {Kennicutt},
  {Lee}, {Dale}, {Engelbracht}, {van Zee}, {Draine}, {Hao}, {Gordon},
  {Moustakas}, {Murphy}, {Regan}, {Begum}, {Block}, {Dalcanton}, {Funes}, {Gil
  de Paz}, {Johnson}, {Sakai}, {Skillman}, {Walter}, {Weisz}, {Williams}, \&
  {Wu}}]{calzetti10}
{Calzetti}, D., {Wu}, S.-Y., {Hong}, S., {Kennicutt}, R.~C., {Lee}, J.~C.,
  {Dale}, D.~A., {Engelbracht}, C.~W., {van Zee}, L., {Draine}, B.~T., {Hao},
  C.-N., {Gordon}, K.~D., {Moustakas}, J., {Murphy}, E.~J., {Regan}, M.,
  {Begum}, A., {Block}, M., {Dalcanton}, J., {Funes}, J., {Gil de Paz}, A.,
  {Johnson}, B., {Sakai}, S., {Skillman}, E., {Walter}, F., {Weisz}, D.,
  {Williams}, B., \& {Wu}, Y. 2010, ApJ, 714, 1256

\bibitem[{{Carilli} \& {Walter}(2013)}]{carilli13}
{Carilli}, C.~L. \& {Walter}, F. 2013, ARA\&A (in press), arXiv:1301:0371

\bibitem[{{Chary} {et~al.}(2005){Chary}, {Stern}, \& {Eisenhardt}}]{chary05}
{Chary}, R.-R., {Stern}, D., \& {Eisenhardt}, P. 2005, ApJ, 635, L5

\bibitem[{{Chung} {et~al.}(2007){Chung}, {van Gorkom}, {Kenney}, \&
  {Vollmer}}]{chung07}
{Chung}, A., {van Gorkom}, J.~H., {Kenney}, J.~D.~P., \& {Vollmer}, B. 2007,
  ApJ, 659, L115

\bibitem[{{Cowie} {et~al.}(2011){Cowie}, {Hu}, \& {Songaila}}]{cowie11}
{Cowie}, L.~L., {Hu}, E.~M., \& {Songaila}, A. 2011, ApJ, 735, L38

\bibitem[{{Crawford} {et~al.}(1985){Crawford}, {Genzel}, {Townes}, \&
  {Watson}}]{crawford85}
{Crawford}, M.~K., {Genzel}, R., {Townes}, C.~H., \& {Watson}, D.~M. 1985, ApJ,
  291, 755

\bibitem[{{Dannerbauer} {et~al.}(2009){Dannerbauer}, {Daddi}, {Riechers},
  {Walter}, {Carilli}, {Dickinson}, {Elbaz}, \& {Morrison}}]{dannerbauer09}
{Dannerbauer}, H., {Daddi}, E., {Riechers}, D.~A., {Walter}, F., {Carilli},
  C.~L., {Dickinson}, M., {Elbaz}, D., \& {Morrison}, G.~E. 2009, ApJ, 698,
  L178

\bibitem[{{De Breuck} {et~al.}(2011){De Breuck}, {Maiolino}, {Caselli},
  {Coppin}, {Hailey-Dunsheath}, \& {Nagao}}]{debreuck11}
{De Breuck}, C., {Maiolino}, R., {Caselli}, P., {Coppin}, K.,
  {Hailey-Dunsheath}, S., \& {Nagao}, T. 2011, A\&A, 530, L8

\bibitem[{{de Looze} {et~al.}(2011){de Looze}, {Baes}, {Bendo}, {Cortese}, \&
  {Fritz}}]{delooze11}
{de Looze}, I., {Baes}, M., {Bendo}, G.~J., {Cortese}, L., \& {Fritz}, J. 2011,
  MNRAS, 416, 2712

\bibitem[{{Fixsen} {et~al.}(1999){Fixsen}, {Bennett}, \& {Mather}}]{fixsen99}
{Fixsen}, D.~J., {Bennett}, C.~L., \& {Mather}, J.~C. 1999, ApJ, 526, 207

\bibitem[{{Gunn} \& {Gott}(1972)}]{gunn72}
{Gunn}, J.~E. \& {Gott}, III, J.~R. 1972, ApJ, 176, 1

\bibitem[{{Hu} {et~al.}(2002a){Hu}, {Cowie}, {McMahon}, {Capak}, {Iwamuro},
  {Kneib}, {Maihara}, \& {Motohara}}]{hu02}
{Hu}, E.~M., {Cowie}, L.~L., {McMahon}, R.~G., {Capak}, P., {Iwamuro}, F.,
  {Kneib}, J.-P., {Maihara}, T., \& {Motohara}, K. 2002a, ApJ, 568, L75

\bibitem[{{Hu} {et~al.}(2002b){Hu}, {Cowie}, {McMahon}, {Capak}, {Iwamuro},
  {Kneib}, {Maihara}, \& {Motohara}}]{hu02err}
{Hu}, E.~M., {Cowie}, L.~L., {McMahon}, R.~G., {Capak}, P., {Iwamuro}, F.,
  {Kneib}, J.-P., {Maihara}, T., \& {Motohara}, K. 2002b, ApJ, 576, L99

\bibitem[{{Kashikawa} {et~al.}(2011){Kashikawa}, {Shimasaku}, {Matsuda},
  {Egami}, {Jiang}, {Nagao}, {Ouchi}, {Malkan}, {Hattori}, {Ota}, {Taniguchi},
  {Okamura}, {Ly}, {Iye}, {Furusawa}, {Shioya}, {Shibuya}, {Ishizaki}, \&
  {Toshikawa}}]{kashikawa11}
{Kashikawa}, N., {Shimasaku}, K., {Matsuda}, Y., {Egami}, E., {Jiang}, L.,
  {Nagao}, T., {Ouchi}, M., {Malkan}, M.~A., {Hattori}, T., {Ota}, K.,
  {Taniguchi}, Y., {Okamura}, S., {Ly}, C., {Iye}, M., {Furusawa}, H.,
  {Shioya}, Y., {Shibuya}, T., {Ishizaki}, Y., \& {Toshikawa}, J. 2011, ApJ,
  734, 119

\bibitem[{{Kennicutt}(1998)}]{kennicutt98}
{Kennicutt}, Jr., R.~C. 1998, ApJ, 498, 541

\bibitem[{{Kneib} {et~al.}(1993){Kneib}, {Mellier}, {Fort}, \&
  {Mathez}}]{kneib93}
{Kneib}, J.~P., {Mellier}, Y., {Fort}, B., \& {Mathez}, G. 1993, A\&A, 273, 367

\bibitem[{{Leitherer} {et~al.}(1999){Leitherer}, {Schaerer}, {Goldader},
  {Gonz{\'a}lez Delgado}, {Robert}, {Kune}, {de Mello}, {Devost}, \&
  {Heckman}}]{leitherer99}
{Leitherer}, C., {Schaerer}, D., {Goldader}, J.~D., {Gonz{\'a}lez Delgado},
  R.~M., {Robert}, C., {Kune}, D.~F., {de Mello}, D.~F., {Devost}, D., \&
  {Heckman}, T.~M. 1999, ApJS, 123, 3

\bibitem[{{Maiolino} {et~al.}(2009){Maiolino}, {Caselli}, {Nagao}, {Walmsley},
  {De Breuck}, \& {Meneghetti}}]{maiolino09}
{Maiolino}, R., {Caselli}, P., {Nagao}, T., {Walmsley}, M., {De Breuck}, C., \&
  {Meneghetti}, M. 2009, A\&A, 500, L1

\bibitem[{{Maiolino} {et~al.}(2005){Maiolino}, {Cox}, {Caselli}, {Beelen},
  {Bertoldi}, {Carilli}, {Kaufman}, {Menten}, {Nagao}, {Omont}, {Wei{\ss}},
  {Walmsley}, \& {Walter}}]{maiolino05}
{Maiolino}, R., {Cox}, P., {Caselli}, P., {Beelen}, A., {Bertoldi}, F.,
  {Carilli}, C.~L., {Kaufman}, M.~J., {Menten}, K.~M., {Nagao}, T., {Omont},
  A., {Wei{\ss}}, A., {Walmsley}, C.~M., \& {Walter}, F. 2005, A\&A, 440, L51

\bibitem[{{Mellier} {et~al.}(1988){Mellier}, {Soucail}, {Fort}, \&
  {Mathez}}]{mellier88}
{Mellier}, Y., {Soucail}, G., {Fort}, B., \& {Mathez}, G. 1988, A\&A, 199, 13

\bibitem[{{Pierini} {et~al.}(2001){Pierini}, {Lequeux}, {Boselli}, {Leech}, \&
  {V{\"o}lk}}]{pierini01}
{Pierini}, D., {Lequeux}, J., {Boselli}, A., {Leech}, K.~J., \& {V{\"o}lk},
  H.~J. 2001, A\&A, 373, 827

\bibitem[{{Richard} {et~al.}(2010){Richard}, {Kneib}, {Limousin}, {Edge}, \&
  {Jullo}}]{richard10}
{Richard}, J., {Kneib}, J.-P., {Limousin}, M., {Edge}, A., \& {Jullo}, E. 2010,
  MNRAS, 402, L44

\bibitem[{{Schaerer} \& {Pell{\'o}}(2005)}]{schaerer05}
{Schaerer}, D. \& {Pell{\'o}}, R. 2005, MNRAS, 362, 1054

\bibitem[{{Shim} \& {Chary}(2013)}]{shim13}
{Shim}, H. \& {Chary}, R.-R. 2013, ApJ, 765, 26

\bibitem[{{Shim} {et~al.}(2011){Shim}, {Chary}, {Dickinson}, {Lin}, {Spinrad},
  {Stern}, \& {Yan}}]{shim11}
{Shim}, H., {Chary}, R.-R., {Dickinson}, M., {Lin}, L., {Spinrad}, H., {Stern},
  D., \& {Yan}, C.-H. 2011, ApJ, 738, 69

\bibitem[{{Solomon} \& {Barrett}(1991)}]{solomon91}
{Solomon}, P.~M. \& {Barrett}, J.~W. 1991, in IAU Symposium, Vol. 146, Dynamics
  of Galaxies and Their Molecular Cloud Distributions, ed. F.~{Combes} \&
  F.~{Casoli}, 235

\bibitem[{{Spergel} {et al.}(2007)}]{spergel07}
{Spergel}, D.~N. {et al.} 2007, ApJS, 170, 377

\bibitem[{{Stacey} {et~al.}(1991){Stacey}, {Geis}, {Genzel}, {Lugten},
  {Poglitsch}, {Sternberg}, \& {Townes}}]{stacey91}
{Stacey}, G.~J., {Geis}, N., {Genzel}, R., {Lugten}, J.~B., {Poglitsch}, A.,
  {Sternberg}, A., \& {Townes}, C.~H. 1991, ApJ, 373, 423

\bibitem[{{Stacey} {et~al.}(2010){Stacey}, {Hailey-Dunsheath}, {Ferkinhoff},
  {Nikola}, {Parshley}, {Benford}, {Staguhn}, \& {Fiolet}}]{stacey10}
{Stacey}, G.~J., {Hailey-Dunsheath}, S., {Ferkinhoff}, C., {Nikola}, T.,
  {Parshley}, S.~C., {Benford}, D.~J., {Staguhn}, J.~G., \& {Fiolet}, N. 2010,
  ApJ, 724, 957

\bibitem[{{Taniguchi} {et al.}(2005)}]{taniguchi05}
{Taniguchi}, Y. {et al.} 2005, PASJ, 57, 165

\bibitem[{{Wagg} {et~al.}(2010){Wagg}, {Carilli}, {Wilner}, {Cox}, {De Breuck},
  {Menten}, {Riechers}, \& {Walter}}]{wagg10}
{Wagg}, J., {Carilli}, C.~L., {Wilner}, D.~J., {Cox}, P., {De Breuck}, C.,
  {Menten}, K., {Riechers}, D.~A., \& {Walter}, F. 2010, A\&A, 519, L1

\bibitem[{{Wagg} \& {Kanekar}(2012)}]{wagg12a}
{Wagg}, J. \& {Kanekar}, N. 2012, ApJ, 751, L24

\bibitem[{{Wagg} {et~al.}(2009){Wagg}, {Kanekar}, \& {Carilli}}]{wagg09}
{Wagg}, J., {Kanekar}, N., \& {Carilli}, C.~L. 2009, ApJ, 697, L33

\bibitem[{{Walter} {et~al.}(2012){Walter}, {Decarli}, {Carilli}, {Riechers},
  {Bertoldi}, {Wei{\ss}}, {Cox}, {Neri}, {Maiolino}, {Ouchi}, {Egami}, \&
  {Nakanishi}}]{walter12a}
{Walter}, F., {Decarli}, R., {Carilli}, C., {Riechers}, D., {Bertoldi}, F.,
  {Wei{\ss}}, A., {Cox}, P., {Neri}, R., {Maiolino}, R., {Ouchi}, M., {Egami},
  E., \& {Nakanishi}, K. 2012, ApJ, 752, 93

\bibitem[{{Walter} {et al.}(2003)}]{walter03}
{Walter}, F. {et al.} 2003, Nature, 424, 406

\bibitem[{{Wang} {et~al.}(2010){Wang}, {Carilli}, {Neri}, {Riechers}, {Wagg},
  {Walter}, {Bertoldi}, {Menten}, {Omont}, {Cox}, \& {Fan}}]{wang10}
{Wang}, R., {Carilli}, C.~L., {Neri}, R., {Riechers}, D.~A., {Wagg}, J.,
  {Walter}, F., {Bertoldi}, F., {Menten}, K.~M., {Omont}, A., {Cox}, P., \&
  {Fan}, X. 2010, ApJ, 714, 699

\bibitem[{{Wang} {et~al.}(2011){Wang}, {Wagg}, {Carilli}, {Walter}, {Riechers}, {Willott},
  {Bertoldi}, {Omont}, {Beelen}, {Cox}, {Strauss}, {Bergeron}, {Forveille}, 
  {Menten}, \& {Fan}}]{wang11}
  {Wang}, R., {Wagg}, J., {Carilli}, C.~L., {Walter}, F., {Riechers}, D.~A., 
  {Willott}, C., {Bertoldi}, F., {Omont}, A., {Beelen}, A., {Cox}, P., 
  {Strauss}, M.~A., {Bergeron}, J., {Forveille}, T., {Menten}, K.~M., \& {Fan}, X.
  2011, ApJ, 739, L34

\bibitem[{{Wolfire} {et~al.}(1995){Wolfire}, {Hollenbach}, {McKee}, {Tielens},
  \& {Bakes}}]{wolfire95}
{Wolfire}, M.~G., {Hollenbach}, D., {McKee}, C.~F., {Tielens}, A.~G.~G.~M., \&
  {Bakes}, E.~L.~O. 1995, ApJ, 443, 152

\end{thebibliography}

\end{document}